\documentclass{ws-p8-50x6-00}

\begin{document}

\title{Studies of reaction dynamics in the Fermi energy domain }

\author{M. Veselsky, G.A. Souliotis, S.J. Yennello}

\address{Cyclotron Institute, Texas A\&M University, 
College Station TX 77843,
USA\\E-mail: veselsky@comp.tamu.edu}

\maketitle

\abstracts{
An overview of recent results on reaction dynamics in
the energy region 20 - 50 A.MeV is given. The results
of the study of projectile multifragmentation using the detector
array FAUST are presented. Reaction mechanism is determined 
and thermodynamical properties of the hot quasiprojectile are investigated. 
Preliminary results on fragment isospin asymmetry obtained using the 
4$\pi$ detector array NIMROD are given. Procedure for selecting 
centrality in two-dimensional multiplicity histograms is described. 
Possibility to extract thermodynamical temperature from 
systematics of isotope ratios is investigated. 
Reaction mechanism leading to production of hot sources 
is discussed. Furthermore, the possibilities for
production of rare isotopes are discussed and
recent experimental results obtained using recoil
separator MARS are presented.           
}

\section{Introduction \label{sec:intr}}

Nucleus-nucleus collisions in the Fermi energy domain 
( projectile energies 20 - 100 \hbox{A MeV} ) exhibit large 
variety of possible scenarios. In very peripheral collisions 
the relative motion of the projectile and target nuclei 
is mostly tangential and only few nucleons are exchanged 
in the region close to the contact of the surfaces. In more 
damped peripheral collisions nuclei still preserve their identity 
and a di-nuclear system is created due to friction forces. 
In such cases a considerable amount of nucleons can be 
exchanged and hot nuclei can be formed. With further 
decrease of the impact parameter, the energy of the radial  
motion increases and violent scenarios can take place 
where essential parts of both nuclei join into one hot nucleus. 
Hot nuclei created in both peripheral and central collisions 
can undergo multifragmentation. The process of multifragmentation 
is of interest for studies of thermodynamic properties 
of nuclear systems. Experimental studies of multifragmentation 
require multidetector systems with large geometric coverage. 
An essential information about the hot system can be collected 
using such devices and detailed studies can 
be carried out. In this proceeding we present results 
of multifragmentation studies using the forward array FAUST 
and the 4$\pi$ detector array NIMROD. Properties of the 
hot emitting source are investigated. From the practical 
point of view, reactions in the Fermi energy 
domain can provide a new approach to production of rare 
beams. In particular the projectile-like nuclei are of interest 
for the production of rare isotopes because of the extensive 
nucleon exchange. We present here the results of studies 
of $^{86}Kr$ fragmentation at 25 \hbox{A MeV}  
using recoil separator MARS along with comparison to results 
of model calculations.  

\section{Projectile multifragmentation \label{sec:projmf}}
 
Projectile fragmentation has traditionally been thought of as a two-step 
reaction with excitation via a peripheral collision with the target 
followed by fragmentation of the projectile. In this framework, 
the influence of the mass 
and charge of the target nucleus on projectile fragmentation is a 
question of interest both with regard to formation of the excited 
quasiprojectile and its subsequent fragmentation. 
A cycle of works \cite{laforest,MVPRC,MVPRCRC,MVPLB} 
on projectile multifragmentation 
of a $^{28}$Si beam in the reaction with $^{112}$Sn and $^{124}$Sn 
targets at 30 and 50 \hbox{A MeV} was carried out recently at the 
Cyclotron Institute of Texas A\&M University.  

The experiment was done with a beam of $^{28}$Si impinging on $\sim  $1
mg/cm$^{2}$ self supporting $^{112,124}$Sn targets. The beam was
delivered at 30 and 50 \hbox{A MeV} by the K500 superconducting cyclotron at
the Cyclotron Institute of Texas A\&M University. The detector array 
FAUST \cite{FAUST} consisting of 68 silicon - CsI(Tl) telescopes covering 
polar angles from 2.3$^{\circ}$ to 33.6$^{\circ}$ in the laboratory system 
was used. The detectors are arranged in five concentric rings. The geometrical 
efficiency is approximately 90$ \%$ for the angle range covered.  
Isotopes of light charged particles and intermediate-mass fragments 
up to a charge of $ Z_{f}=5 $ were identified. 
Additional particle telescopes at angles between 42.5-82.5$^{\circ }$ and 
123-147$^{\circ }$ complemented the forward array in the setup. 
Details of the experimental 
procedure and detector calibration can be found in ref. \cite{laforest}. 

The study \cite{MVPRC} was restricted to events where 
all emitted fragments were isotopically identified ( $ Z_{f}<5 $ ). We assumed 
that such events detected in the FAUST detector array originate predominantly 
from the deexcitation of the quasiprojectile ( or projectile-like 
source ). This assumption was supported by the Gaussian shape of the 
quasiprojectile velocity distributions and by low multiplicities 
of coincident charged particles at backward angles \cite{MVPRC}. Considerable 
asymmetry of the projectile-target system leads already in peripheral 
collisions to production of highly excited light quasiprojectiles while 
the target nucleus remains much colder and de-excites mostly via emission 
of neutrons. The influence of pre-equilibrium emission was found \cite{MVPRC} 
to be weak and could not influence the reaction scenario dramatically.  
The total charge of the reconstructed quasiprojectile ( QP ) 
was restricted to the values near the projectile charge ( $ Z_{QP}=12-15 $ ). 
This very selective data contains information on fragmentation 
of highly excited projectile-like prefragments, and thus can be used to  
study the mechanism of dissipation of the kinetic energy of relative 
motion into thermal degrees of freedom.

Useful experimental information about the nucleon exchange rate can be found
in the events where the charge of the reconstructed quasiprojectile is equal 
to the charge of incident beam ( $ Z_{QP}=14 $ ). In this case, isospin 
equilibration may only occur by the transfer of neutrons, as the number 
of transferred neutrons is the only available isospin degree of freedom 
of the system. Since the neutron number of the reconstructed 
quasiprojectile is just a sum of neutrons bound in the fragments 
with non-zero charge, we defined a principal 
neutron exchange observable as the mass change. 
Subtracting the sum of the neutrons bound in the 
detected fragments from the neutron number of the beam gives 

\begin{equation}
 \Delta A = N_{proj}-\sum_{f} N_{f} 
\label{dela}
\end{equation}

where $N_{proj} =$ 14 for $ ^{28} $Si beam. 

\begin{figure}[ht]
%\figurebox{20pc}{15pc}{\epsfbox{figure1.eps} }
% {figure1.eps} % to have a box alone
\epsfxsize=18pc % will enlarge or reduce the postscript figures based on the xsize
%\epsfbox{xxx.eps} % postscript image file name
\centerline {\epsfbox{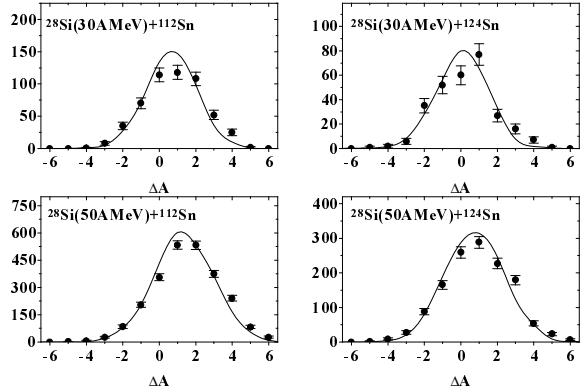} }% postscript image file name
\caption{ Experimental ( solid circles ) and 
simulated ( solid lines ) mass change distributions 
for the fully isotopically resolved quasiprojectiles with $ Z_{QP}=14 $ 
\protect\cite{MVPRC}. \label{fig:fig1}}
\end{figure}

Experimental mass change distributions for both 
projectile energies and target isotopes are shown in Fig. \ref{fig:fig1} 
as circles. 
When looking at the shapes of the observed $ \Delta A$ distributions, 
it is apparent that they are almost identical for different targets 
at the same projectile energy and are close to Gaussians. 
For both projectile energies the mean value of the 
mass change is larger for the reaction with the $^{112}$Sn 
target by a little more than half a unit  
( 0.60 for 30 \hbox{A MeV} and 0.65 for 50 \hbox{A MeV} ). 
The experimental distributions of the mass change are compared to 
simulated ones for fully isotopically resolved events with $ Z_{QP}=14 $ 
( solid lines ). 
The basic assumption of the simulation is the possibility 
to decompose the collision into two stages. 
In the early stage of the collision hot quasiprojectiles are created 
which then deexcite by the statistical decay. To describe the 
production of excited quasiprojectiles we used the Monte Carlo code 
of Tassan-Got et al. \cite{TG}. This code implements a 
version of the model of deep inelastic transfer suitable for Monte Carlo 
simulations. 
De-excitation of the highly excited quasiprojectile was simulated using 
the statistical model of multifragmentation ( SMM ) \cite{SMM}.
Macrocanonical partitions of fragments were generated 
for individual events. 
The agreement of the experimental and simulated distributions of 
the mass change is reasonable. This implies that both the nucleon 
exchange and de-excitation are described adequately. To further 
justify such a conclusion, an apparent charged particle excitation energy 
of the quasiprojectile was reconstructed for each projectile fragmentation 
event from the energy balance in the center of mass frame 
of the quasiprojectile. Thus 

\begin{equation}
 E_{app}^{*}=\sum _{f}(T_{f}^{QP}+\Delta m_{f})-\Delta m_{QP} \hbox{ , }
\label{eq:exqp}
\end{equation}

where $ T_{f}^{QP} $ is the kinetic energy of the fragment in the 
reference frame of the quasiprojectile and $ \Delta m_{f} $ 
and $ \Delta m_{QP} $ are the mass
excesses of the fragment and quasiprojectile, respectively. 

\begin{figure}[ht]
%\figurebox{20pc}{15pc}{\epsfbox{figure1.eps} }
% {figure1.eps} % to have a box alone
\epsfxsize=18pc % will enlarge or reduce the postscript figures based on the xsize
%\epsfbox{xxx.eps} % postscript image file name
\centerline {\epsfbox{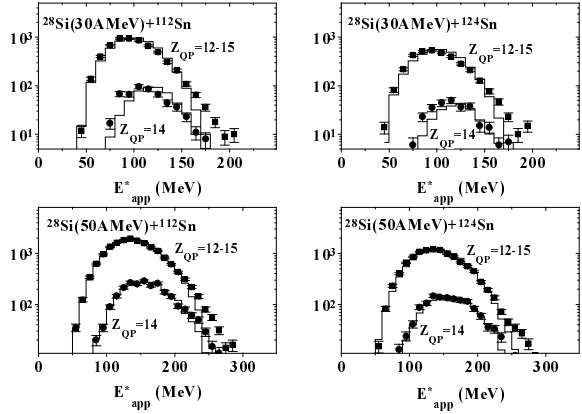} }% postscript image file name
\caption{ Distributions of apparent excitation energies of 
the quasiprojectiles. Symbols mean experimental distributions of  
the set of isotopically resolved quasiprojectiles  
with $ Z_{QP}=14 $ ( solid circles ) and $ Z_{QP}=12-15 $ ( solid squares ) 
\protect\cite{MVPRC}. 
Solid histograms mean simulated distributions.  \label{fig:fig2}}
\end{figure}

The distributions of apparent quasiprojectile excitation energies 
reconstructed from fully isotopically resolved events are shown 
in Fig. \ref{fig:fig2}. The reconstructed distributions 
for multifragmentation events with $ Z_{QP}=14 $ are represented 
as circles. Squares represent a broader set of events with $ Z_{QP}=12-15 $. 
In Fig. \ref{fig:fig2} are also shown the simulated distributions 
of apparent quasiprojectile excitation energies for 
both $ Z_{QP}=14 $ and $ Z_{QP}=12-15 $ ( solid histograms ). 
The simulated data have been 
normalized to the sum of experimental events with $ Z_{QP}=12-15 $. The 
agreement of the simulated and experimental apparent quasiprojectile 
excitation energy
distributions with both $ Z_{QP}=12-15 $ and $ Z_{QP}=14 $ is 
quite good. 
The onset of multifragmentation into channels with 
$ Z_{f} \leq 5 $ in the low energy part is described with good precision 
for both sets of data $ Z_{QP}=12-15 $ and $ Z_{QP}=14 $. 

\begin{figure}[ht]
%\figurebox{20pc}{15pc}{\epsfbox{figure1.eps} }
% {figure1.eps} % to have a box alone
\epsfxsize=20pc % will enlarge or reduce the postscript figures based on the xsize
%\epsfbox{xxx.eps} % postscript image file name
\centerline {\epsfbox{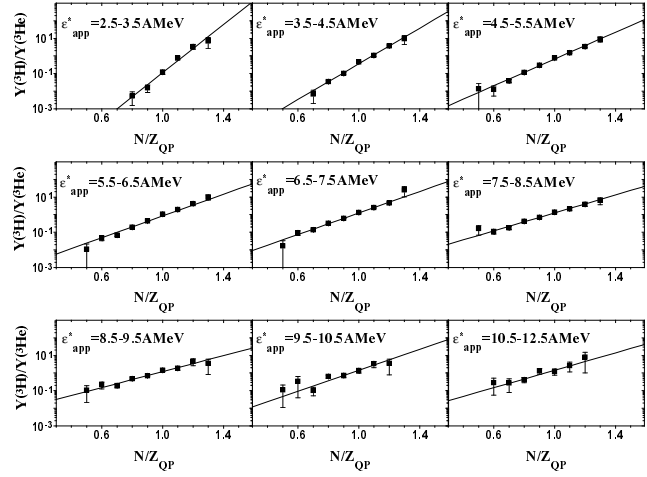} }% postscript image file name
\caption{ 
Dependence of the yield ratio Y($^{3}$H)/Y($^{3}$He) on N/Z ratio of the 
isotopically resolved quasiprojectiles with $Z_{QP}=12-15$ for 
nine bins of $\epsilon^{*}_{app}$ \protect\cite{MVPLB}. 
\label{fig:fig3}}
\end{figure}

The set of data obtained here allows 
further study of thermal quasiprojectile multifragmentation, especially 
the study of influence of the quasiprojectile isospin asymmetry on properties 
of the fragmenting system. 
This data is of specific interest because the isospin asymmetry of the system 
which actually undergoes multifragmentation is known with good precision.  
Fig. \ref{fig:fig3} shows dependences of the isobaric yield 
ratio Y($^{3}$H)/Y($^{3}$He) on $N/Z_{QP}$ for nine bins of the apparent 
excitation energy per mass unit of the quasiprojectile 
( $\epsilon^{*}_{app}$ ). 
The data for both targets and projectile energies were combined 
to increase the statistics. 
This is possible because for a given excitation energy bin, the 
Y($^{3}$H)/Y($^{3}$He) dependences agree within 
the statistical deviations. The experimental data are represented as squares 
and the lines show the fits. 
As one can see on Fig. \ref{fig:fig3}, linearity is the overall feature 
of the logarithmic plots of the Y($^{3}$H)/Y($^{3}$He) ratio in all 
excitation energy bins and is especially significant in the excitation 
energy bins with high statistics. The slopes are steepest at low 
excitation energies and become flatter with increasing excitation energy. 

\begin{figure}[ht]
%\figurebox{20pc}{15pc}{\epsfbox{figure1.eps} }
% {figure1.eps} % to have a box alone
\epsfxsize=15pc % will enlarge or reduce the postscript figures based on the xsize
%\epsfbox{xxx.eps} % postscript image file name
\centerline {\epsfbox{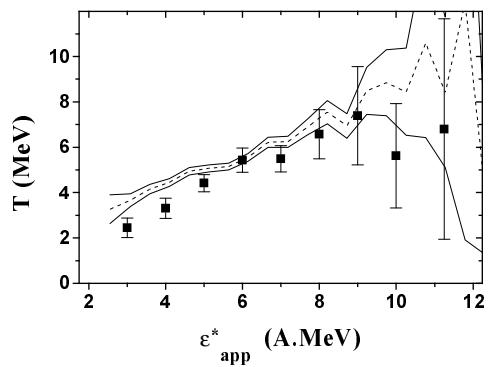} }% postscript image file name
\caption{ 
Dependence of the temperature $ T $, 
determined from the dependence of the yield ratio 
Y($^{3}$H)/Y($^{3}$He) on the N/Z ratio of 
the quasiprojectile, on $\epsilon^{*}_{app}$ 
\hbox{( solid squares )} \protect\cite{MVPLB}. Dashed and solid lines 
indicate the values and the statistical deviations of the double 
isotope ratio thermometer  \hbox{d,t/$^{3}$He,$^{4}$He}.  
\label{fig:fig4}}
\end{figure}

Within the approximation described in detail in \cite{MVPLB}, the expression 
for the isobaric ratio Y($^{3}$H)/Y($^{3}$He) will be 

\begin{equation}
\ln(Y(^{3}\hbox{H})/Y(^{3}\hbox{He})) = \ln(K(T)) + (\mu _{n}-\mu _{p})/T 
\label{eq:r3ln}
\end{equation}

\noindent 
where $ K(T) $ is a proportionality factor dependent on the temperature 
but independent of the N/Z ratio of the fragmenting system. 
The difference of chemical potentials for neutrons and protons 
$\mu _{n}-\mu _{p}$ was approximated by the difference of proton 
and neutron separation energies in the ground state $ S_{p}-S_{n} $. 
Experimental mass excesses \cite{wapstra} have been used for 
the evaluation. The dependence of $ S_{p}-S_{n} $ on the N/Z ratio of 
the quasiprojectile can be considered linear with good precision. 
Formula \ref{eq:r3ln} allows a determination not only of the temperature 
( as a ratio of the slopes ) but also of ln($K(T)$) from the comparison 
of the zero order coefficients \hbox{( constants )} using the extracted 
temperature. The resulting values of $ T $ are given in Fig. \ref{fig:fig4} 
for the different quasiprojectile excitation energy bins. 

The dependence of the temperature on the excitation energy ( caloric 
curve ), given in Fig. \ref{fig:fig4} ( solid squares ), 
is compared to the experimental caloric curve 
obtained for the same set of data by the double isotope 
ratio method for the thermometer \hbox{d,t/$^{3}$He,$^{4}$He} 
( dashed line indicates 
the double isotope ratio temperature and solid lines indicate 
statistical \hbox{errors ).} 
The agreement between the two plots is 
reasonable. Both methods give the value of the temperature between 5 and 7 MeV 
for the apparent excitation energies above 5 \hbox{A MeV}. 
The agreement between temperatures determined by the two 
different methods shows that the assumptions made for $ \mu _{n}-\mu _{p} $ 
reflect the physical trends which take place in the freeze-out configuration. 
Since the multiplicity of the charged fragments decreases 
with decreasing $\epsilon^{*}_{app}$, the decrease of the determined 
temperatures in the lowest bins of $\epsilon^{*}_{app}$ may 
be a signature of the onset of a low energy deexcitation mode 
where several neutrons and light charged particles are emitted prior 
to the breakup of the partially cooled residue into two massive fragments 
( such a mass distribution is observed in the channels with 3 and 4 charged 
fragments ). The temperatures determined thus become mean values 
representing the range of temperatures at which hydrogen and helium 
isotopes are emitted during the deexcitation cascade. 

\section{Isospin asymmetry of reaction products originating 
from projectile-target systems with different N/Z \label{sec:isasm}}

With increasing violence of the nucleus-nucleus collision 
a hot composite system can be created. Such a source can 
consist of parts of both the projectile and target. The excitation 
energy of such a source is high enough for emission of 
intermediate mass fragments ( IMF ). Another possibility for 
emission of intermediate mass fragments is a dynamical 
scenario where a neck is formed and IMFs are remnants 
of its rupture. Information on both mass and charge of 
emitted IMFs can be crucial for the determination of the emission 
mechanism and thus properties of the emitting source.  

Four reactions of $^{124}$Sn,$^{124}$Xe beams with $^{112,124}$Sn 
targets have been studied at 28 \hbox{A MeV} \cite{MVNim} at 
the Cyclotron Institute of Texas A\&M University 
using the 4$\pi$ multi-detector array NIMROD \cite{Nimrod}. 
The aim of the experiment was to study the interplay of isospin
degree of freedom with reaction dynamics of the projectile-target system. 
NIMROD is a 4$\pi$ neutron and charged particle
detection system. Neutrons are detected using a liquid scintillator which
is contained in vessels around the target. Charged
particles are detected using 96 charged particle
detection modules in 12 rings. A typical detection module consists 
of a gas ionization chamber and one or two CsI(Tl) detectors. 
In each ring there are several modules where one or two silicon detectors 
are placed between the ionization chamber and CsI(Tl) detector. 
NIMROD has nearly complete coverage for both charged particles and neutrons 
as well as selective coverage for isotopically resolved charged
particles up to oxygen. An isospin asymmetry of the reaction products 
can be examined with respect to the observables
characterizing the dynamical evolution of different 
projectile-target systems. The systems studied have 
significantly different values of initial
isospin asymmetries of both projectile and target. 

\begin{figure}[ht]
%\figurebox{20pc}{15pc}{\epsfbox{figure1.eps} }
% {figure1.eps} % to have a box alone
\epsfxsize=19pc % will enlarge or reduce the postscript figures based on the xsize
%\epsfbox{xxx.eps} % postscript image file name
\centerline {\epsfbox{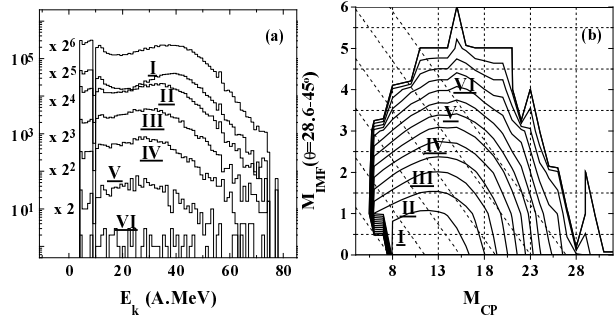} }% postscript image file name
\caption{ (a) - Inclusive spectrum of $\alpha$-particles ( top ) 
from reaction $^{124}$Xe+$^{124}$Sn and spectra for 6 centrality 
bins ordered from top to bottom by increasing centrality \protect\cite{MVNim}. 
(b) - Centrality cuts in the M$_{IMF}$ vs. M$_{CP}$ histogram. 
Skewed lines indicate centrality cuts used in (a), horizontal 
and vertical lines indicate one-dimensional cuts in 
M$_{IMF}$ and M$_{CP}$ histograms. \label{fig:fig5}}
\end{figure}

Newly developed calibration methods \cite{MVCal} have been employed 
in data analysis. The calibration coefficients are obtained 
from a minimization procedure where three isotope lines assigned 
in the experimental two-dimensional spectra are fitted to the 
calculated energy losses in the telescope. A functional for 
dependence of energy on light output in the CsI(Tl) detector 
was taken from ref. \cite{ltgcsi}. An example of calibrated 
$\alpha$-spectra \cite{MVNim} with different centralities is shown 
in Fig. \ref{fig:fig5}a. 
The spectra were collected at angles 11.0-14.9$^{\circ}$ in one 
of the Si-Si-CsI(Tl) telescopes. Both inclusive spectrum ( top one ) and 
spectra collected in several centrality 
bins ( centrality increases from top to bottom, multiplication 
coefficients of spectra differ by factor of two ). Centrality cuts 
have been made in the two-dimensional histogram of IMF multiplicities 
( M$_{IMF}$ ) vs multiplicities of charged particles ( M$_{CP}$ ) 
by parallel cuts chosen so that the most central cut selects the events 
along the line connecting the events with highest multiplicities of charged 
particles and highest IMF multiplicities ( see Fig. \ref{fig:fig5}b ). 
Such a selection unifies both criteria and adds events of analogous 
centrality where the content of IMFs and charged particles is between both 
extremes. Event shape analysis was carried out and showed analogous trends 
in terms of sphericity and coplanarity. With increasing centrality 
the M$_{IMF}$ criterion selects the values of flow angle still closer 
to 90$^{\circ}$, while M$_{CP}$ criterion selects still lower flow angles. 
The unified criterion selects events with slowly changing flow angles between 
the former two. The statistics in the most central cut increased by two orders 
of magnitude. Fig. \ref{fig:fig5}a shows that also the shapes of spectra change 
gradually with increasing centrality. 

\begin{figure}[ht]
%\figurebox{20pc}{15pc}{\epsfbox{figure1.eps} }
% {figure1.eps} % to have a box alone
\epsfxsize=20pc % will enlarge or reduce the postscript figures based on the xsize
%\epsfbox{xxx.eps} % postscript image file name
\centerline {\epsfbox{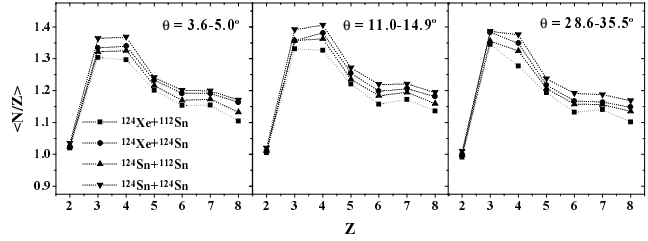} }% postscript image file name
\caption{ Mean N/Z-ratios of emitted LCPs and IMFs with Z=2-8 from 
four reactions obtained using the inclusive data from three Si-Si-CsI(Tl) 
telescopes positioned at different angles \protect\cite{MVNim}. 
\label{fig:fig6}}
\end{figure}

In Fig. \ref{fig:fig6} are given the mean N/Z-ratios of emitted LCPs and IMFs 
with Z=2-8 from four reactions obtained using the inclusive data 
from three Si-Si-CsI(Tl) telescopes positioned at different angles 
\cite{MVNim}. 
The overall dependence is similar at all angle ranges. 
He-isotopes ( Z=2 ) are dominated by $\alpha$-particles. 
Mean N/Z ratios are highest for Li- ( Z=3 ) and Be-isotopes ( Z=4 ) 
and decrease gradually with increasing atomic number of IMFs. 
When comparing mean N/Z ratios from four reactions of 
$^{124}$Sn,$^{124}$Xe beams with $^{112,124}$Sn targets 
two characteristic patterns can be distinguished. At forward 
angles 3.6-5.0$^{\circ}$ mean N/Z ratios of Li- and Be-isotopes appear 
to track with the isospin asymmetry of the whole projectile-target 
system what creates a typical 1-2-1 pattern. On the other hand 
the isotopes with Z=5-8 appear to track with the isospin asymmetry 
of the target nucleus what creates a 2-2 pattern. At angles 11.0-14.9$^{\circ}$ 
the 1-2-1 pattern can be identified for Li- and Be-isotopes 
while for heavier fragments the pattern vary for different atomic numbers. 
At angles 28.6-35.5$^{\circ}$ one can recognize the 2-2 pattern 
for Li-isotopes while at heavier fragments the 1-2-1 pattern dominates. 
In principle, one can identify the 1-2-1 pattern to the origin of fragments 
emitted from hot composite source and 2-2 pattern to the emission from 
quasiprojectile. The fragments with Z=5-8 at forward angles can be 
possibly attributed to binary de-excitation channels of the moderately 
excited quasiprojectile while the 2-2 pattern of Li-isotopes at the most 
central angle range may suggest either backward emission from the 
quasiprojectile or emission from quasitarget. Similar isospin asymmetry 
patterns ( 1-2-1 and 2-2 ) can be observed also in the isotope yield 
distributions. More detailed studies 
including also constraints on different centralities are now in progress. 
In any case the data demonstrates that the isospin asymmetry of emitted 
fragments is a sensitive probe of the reaction dynamics. 

\begin{figure}[ht]
%\figurebox{20pc}{15pc}{\epsfbox{figure1.eps} }
% {figure1.eps} % to have a box alone
\epsfxsize=20pc % will enlarge or reduce the postscript figures based on the xsize
%\epsfbox{xxx.eps} % postscript image file name
\centerline {\epsfbox{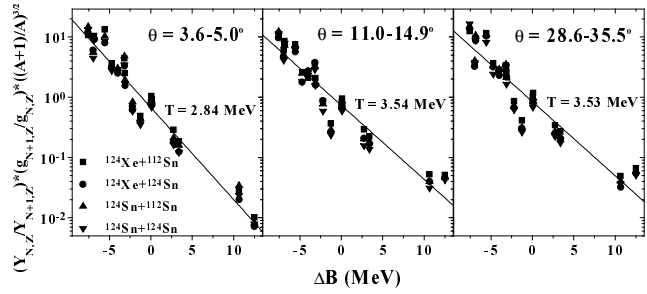} }% postscript image file name
\caption{ Systematics of corrected isotopic ratios $Y_{N,Z}/Y_{N+1,Z}$ from 
four reactions plotted against the difference of binding energies 
 \protect\cite{MVNim}. \label{fig:fig7}}
\end{figure}

Fig. \ref{fig:fig7} shows the systematics of corrected isotopic ratios 
$Y_{N,Z}/Y_{N+1,Z}$ from the four reactions plotted against the difference 
of binding energies \cite{MVNim}. 
The inclusive data from the three Si-Si-CsI(Tl) telescopes positioned 
at different angles was used. Isotopic ratios were corrected 
to g.s. spin and mass in order to investigate eventually the possibility 
to estimate the temperature at which fragments have been emitted. 
For the macrocanonical ensemble the quantity shown in Fig. \ref{fig:fig7} 
becomes proportional to $e^{-\frac{\Delta B}{T}}$ when assuming 
that the difference of free energies can be approximated by difference 
of binding energies. The experimental systematics given in Fig. \ref{fig:fig7} 
appears to follow a similar trend. There are no significant differences 
for data from different reactions. Apparent temperature at given angles 
was estimated using exponential fits to data ( see Fig. \ref{fig:fig7} ). It 
is lowest at forward angles and appears to saturate at more central 
angles. It is one of the goals of further analysis to obtain similar 
information for subsets of data with well defined centrality 
and excitation energy and to carry out comparison to various double isotope 
ratio thermometers.  

\section{Production mechanism of the hot source \label{sec:prodmc}}

In order to create a hot source, a considerable part 
of the kinetic energy should be damped into thermal 
degrees of freedom. Hot source can be created both 
in damped peripheral and violent collisions. In damped peripheral 
collisions projectile and target nuclei preserve their identity 
and a di-nuclear system is created due to friction forces. 
In such cases a considerable amount of nucleons can be 
exchanged what leads to dissipation of a significant part 
of the available energy into heat. As demonstrated in section \ref{sec:projmf} 
the model of deep inelastic transfer combined with an appropriate 
de-excitation model can reproduce the experimental data very precisely. 
This will be further demonstrated in the next section. 
With decrease of the impact parameter the energy of the radial  
motion increases and violent scenarios can take place 
where essential parts of both nuclei join into one hot nucleus. 
A hybrid model for such a collisions was developed by one of us 
recently \cite{MVMod}. In order to describe the reaction dynamics consistently, 
pre-equilibrium emission was taken into account phenomenologically. 
The hot source is created by incomplete fusion of the participant 
zone with one of the spectators. The other spectator forms a relatively 
cold source. Motion along classical Coulomb trajectories is assumed. 
The results of the model calculation were compared to wide range 
of experimental data and a consistent description was obtained \cite{MVMod}. 

\begin{figure}[ht]
%\figurebox{20pc}{15pc}{\epsfbox{figure1.eps} }
% {figure1.eps} % to have a box alone
\epsfxsize=20pc % will enlarge or reduce the postscript figures based on the xsize
%\epsfbox{xxx.eps} % postscript image file name
\centerline {\epsfbox{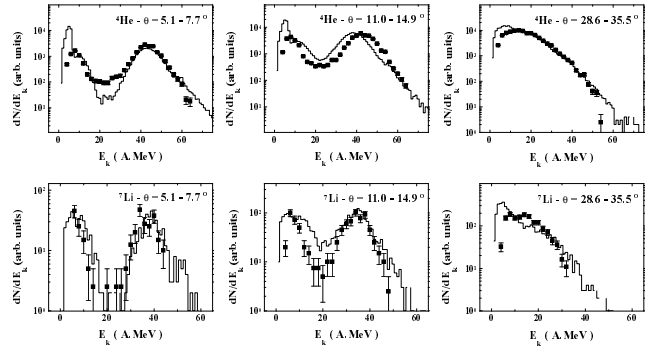} }% postscript image file name
\caption{ 
Experimental spectra ( symbols ) of $\alpha$-particles and 
$^{7}$Li-fragments obtained using NIMROD detector in the reaction 
$^{124}$Sn(28AMeV)+$^{28}$Si at three angles \protect\cite{MVNim} compared 
to the results of model calculation ( histograms ). 
\label{fig:fig8} }
\end{figure}

The capabilities of the model are illustrated 
in Fig. \ref{fig:fig8} where experimental spectra of $\alpha$-particles and 
$^{7}$Li-fragments obtained using the NIMROD detector in the reaction 
$^{124}$Sn(28AMeV)+$^{28}$Si at three angles \cite{MVNim} are compared 
to the results of model calculation. The reaction studied is inverse of 
the reactions studied in section \ref{sec:projmf} and one can expect 
that the high energy IMFs emitted at forward angles originate from 
violent collisions. Nucleon exchange usually leads to the close to equal 
excitation energy sharing what in mass asymmetric systems leads to 
the very hot light nucleus while the heavy one remains cold. 
The excitation energy high enough for emission of IMFs can be 
concentrated in the hot heavy composite source created in violent collisions. 
The SMM code \cite{SMM} was used for the de-excitation of both the 
hot heavy source and the cold target remnant. No mutual Coulomb 
interaction of de-exciting sources was assumed in the calculation. 
The same normalization of the simulated data 
is used for both fragments and all angle ranges. The model calculation 
describes the essential features of the experimental spectra at all angles. 
According to the calculations the dips in the spectra at forward angles 
correspond to the velocity of emitting source. The remaining discrepancies 
can be possibly attributed to the mutual Coulomb interaction between 
the hot and cold sources which was not taken into account. 
In order to account for mutual Coulomb interaction between sources an 
additional assumptions concerning their distance should be introduced. 
Further investigations focused on this topic are in progress. 

\section{Production of neutron-rich nuclides using $^{86}$Kr beam 
 at projectile energy 25 \hbox{A MeV} \label{sec:prodkrni}}

As already pointed out, peripheral reactions between massive nuclei at the 
Fermi energy domain involve considerable exchange of nucleons. As a result,
rare isotopes  ( either proton-rich or neutron-rich ) can be produced with 
large cross sections.   
This possibility, in regards to the production of neutron rich nuclides 
was recently explored in the reaction of neutron-rich $^{86}$Kr-beam  with 
a neutron-rich $^{64}$Ni-target.

In a recent study \cite{GSPaper}, a  25 \hbox{A MeV} $^{86}$Kr$^{22+}$ beam 
from 
the K500 superconducting cyclotron, with a typical current of $\sim$1 pnA, 
interacted with a $^{64}$Ni target of  thickness 4 mg/cm$^{2}$.
The reaction products were analyzed with  the MARS spectrometer \cite{MARS}.
The primary beam struck the target at 0$^{o}$ relative to the optical 
axis of the spectrometer.  The direct beam was collected in  a small square 
Faraday cup
( blocking the angular range 0-1$^{o}$ ),
while the fragments were accepted in the angular range 1.0--2.7$^{o}$. 
MARS optics \cite{MARS} provides one  intermediate dispersive  image and a 
final achromatic image ( focal plane ). At the focal plane, 
the fragments were collected in a large area ( 5$\times$5 cm ) three-element 
( $\Delta $E$_{1}$,  $\Delta $E$_{2}$, E ) Si detector telescope. 
The $\Delta$E$_{1}$ detector was a position-sensitive Si strip detector 
of 63 $\mu$m 
thickness whereas the $\Delta$E$_{2}$ and the E detector were  
single-element Si detectors of
150 and  950 $\mu$m, respectively.
Time of flight was measured between two PPACs ( parallel plate avalanche 
counters )
positioned at the dispersive image and at the focal plane, respectively, 
and separated by a distance of 13.2  m. 
The PPAC at the dispersive image was also  X--Y  position sensitive  and  
used  to record 
the position of the reaction products. The horizontal position, along with NMR
measurements of the field of the MARS first dipole, 
was used to determine the magnetic rigidity $B\rho $ of the particles. 
Thus the reaction products were characterized by an event-by-event measurement
of dE/dx, E, time of flight, and magnetic rigidity. 
The response of the spectrometer/detector system 
to ions of known atomic number Z, mass number A, ionic charge q and 
velocity was calibrated using low intensity primary beams of 
$^{40}$Ar and  $^{86}$Kr at 25 \hbox{A MeV}. 

\begin{figure}[htbp]
\epsfxsize=15pc % will enlarge or reduce the postscript figures based on the xsize
%\epsfbox{xxx.eps} % postscript image file name
\centerline {\epsfbox{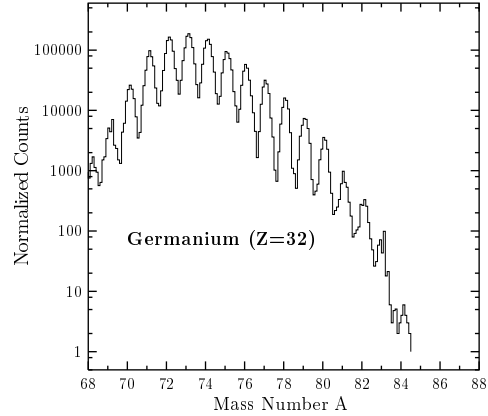} }% postscript image file name
\caption{ Mass histogram of Germanium (Z=32) isotopes \protect\cite{GSPaper}. }
\label{fig:fig9}
\end{figure}

The determination of the atomic number Z was based on the energy loss of the 
particles in the first $\Delta E$ detector and their velocity and is described 
in 
more detail in \cite{GSPaper}. The  Z  resolution was  0.5 units ( FWHM ) for  
near-projectile 
fragments. 
The ionic charge $q$ of the particles entering MARS  was obtained from
the total energy E$_{tot}$=$\Delta$E$_1$+$\Delta$E$_2$+E, the velocity and 
the magnetic rigidity
according to the expression: 
\begin{equation}
q=\frac{3.107}{931.5}\frac{E_{tot}}{B\rho (\gamma -1)}\beta \gamma
\label{q_eqn}
\end{equation}
where E$_{tot}$ is in MeV, B$\rho $ in Tm, $\beta =\upsilon /c$ and $\gamma
=1/(1-\beta ^2)^{\frac 12}$. 
The measurement of the ionic charge q had a resolution of 0.4 units ( FWHM ).
Since the ionic charge must be an integer, we assigned integer
values of q for each event by putting windows ( $\Delta q=0.4$ ) 
on each peak of the q spectrum. 
Using the magnetic rigidity and velocity measurement, the mass-to-charge 
A/q ratio  of each ion was obtained from the expression: 
\begin{equation}
A/q = \frac{B\rho }{3.107\beta \gamma }  \label{Aq_eqn}
\end{equation}
Now, combining the q determination with the A/q measurement, the mass A
was obtained as:
\begin{equation}
A = q_{int} \times A/q  \label{A_eqn}
\end{equation}
( q$_{int}$ is the integer ionic charge determined as above ) with an 
overall resolution 
( FWHM ) of about 0.6 A unit ( See Fig. \ref{fig:fig9} ).

\begin{figure}[htbp]
\epsfxsize=15pc % will enlarge or reduce the postscript figures based on the xsize
%\epsfbox{xxx.eps} % postscript image file name
\centerline {\epsfbox{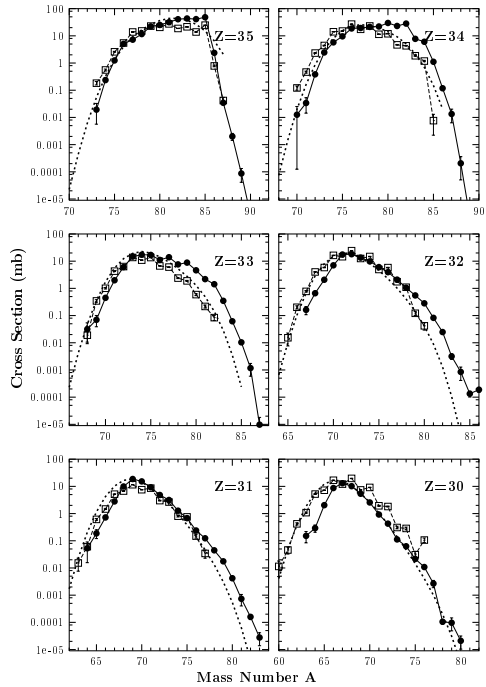} }% postscript image file name
\caption{  Mass distributions of several elements from the reaction of 
           25 \hbox{A MeV} $^{86}$Kr with $^{64}$Ni \protect\cite{GSPaper}. 
	   The present  data
           are shown by full symbols. Open symbols are simulations according 
	   to 
           DIT/GEMINI and the dashed line is from the high-energy 
	   parametrization EPAX 
           ( see text ).}
\label{fig:fig10}
\end{figure}

Combination and  normalization of the data at the various magnetic
rigidity settings of the spectrometer, and summation over all ionic 
charge states
( with corrections applied for missing charge states ),
provided fragment distributions with respect to Z, A  and velocity.
Fig. \ref{fig:fig9} shows the mass spectrum of Z=32 isotopes 
in full resolution.
Results on the mass distributions ( cross sections ) of several elements 
are shown in Fig. \ref{fig:fig10} ( solid points ) 
and are compared to reaction  simulations appropriate for this energy regime. 

The simulations of the present reaction involve the  deep inelastic 
transfer code 
of Tassan-Got \cite{TG} for the primary interaction stage.
Following the creation of the primary fragments,
the statistical de-excitation of the excited primary fragments was simulated
using the code GEMINI \cite{GEMINI}.  
Each partial wave distribution was appropriately weighted 
% by $\pi$$\lambdabar$$^2$ 
and combined to give the overall fragment Z, A ( and velocity ) distributions.
In Fig. \ref{fig:fig10}, the mass distributions for elements Z=30--35, 
calculated by this model, are shown as open squares.
The heavy dashed lines are  predictions of the EPAX parametrization \cite{EPAX}
of relativistic fragmentation cross sections and is plotted here 
for comparison. ( Note that, in high-energy fragmentation, 
nucleon-pickup products are not produced--or, at best, are highly  suppressed
compared to lower energy peripheral collisions ).
As we see in Fig. \ref{fig:fig10}, neutron-rich nuclides are produced 
in substantial yields.
For near projectile elements, an enhancement in the production is observed 
that can not be described by the simulations. This enhancement takes 
place at masses close to the beam and thus in the very peripheral reactions 
where the nucleon exchange can be restricted to neutron-rich surface 
region of the target nucleus. Further investigation of  this finding 
is currently underway. 

\begin{figure}[tbph]
\epsfxsize=15pc % will enlarge or reduce the postscript figures based on the xsize
%\epsfbox{xxx.eps} % postscript image file name
\centerline {\epsfbox{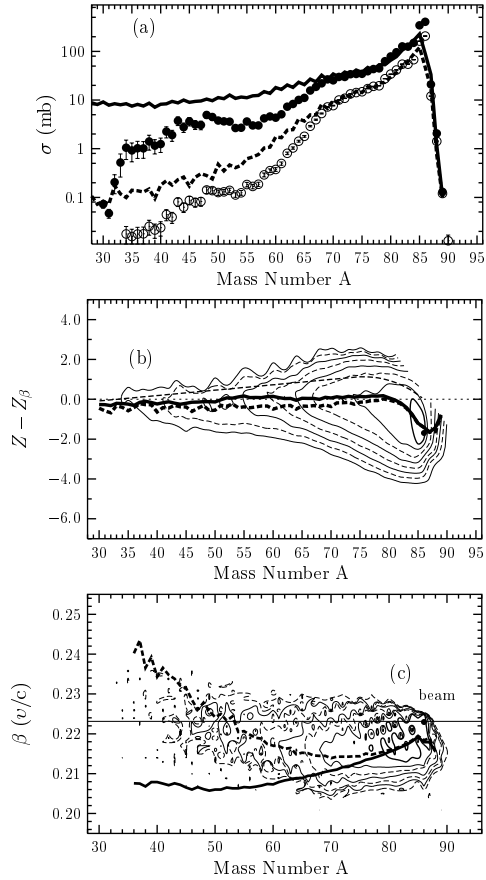} }% postscript image file name
\caption{ 
    Fragment distributions for the reaction of 
    25 \hbox{A MeV} $^{86}$Kr  with  $^{64}$Ni \protect\cite{GSPaper}.
(a) - isobaric yield distribution. The data are shown as solid points. 
    The full line is the result of DIT/GEMINI ( see text ). 
    The dashed  line is  the result of the  same calculation as the full line, 
    but with a cut corresponding  to the angular and momentum acceptance 
    of the spectrometer.  
(b) - yield distributions 
    as a function of Z ( relative to the line of $\protect\beta$ stability,
    Z$_{\protect\beta}$ ) and A.
    Highest yield contours are plotted with thicker lines. Successive contours
    correspond to a decrease  of the yield by a factor of 2.
    The calculated values from DIT/GEMINI are shown as i) thick full line:
    without acceptance cut and, ii) thick dashed line: with acceptance cut.
    Thin dashed line: EPAX parametrization. 
(c) - velocity vs. mass distributions.
  Data are shown as contours as in (b). The thick lines are as in (b).
    The  horizontal full line represents the  beam velocity.
   }
\label{fig:fig11}

\end{figure}

In Fig. \ref{fig:fig11}, the gross features of the distributions are described. 
In Fig. \ref{fig:fig11}a, 
the mass yield curve is presented. The measured data are given as open symbols.
The result of the DIT/GEMINI calculation, filtered by the spectrometer angular 
and momentum acceptance is given by the dashed line, whereas the full line 
gives the total ( unfiltered ) yield. Using the ratio of filtered to unfiltered
calculated yield, correction factors for the acceptance of the spectrometer
were obtained as a function of mass and were applied to the measured yield data 
to obtain the total yield, given by the full symbols in the figure. 
These correction factors were also  employed to obtain total isotope production 
cross sections ( as e.g. given in Fig. \ref{fig:fig10} ) from the measured 
yields. In Fig. \ref{fig:fig11}b,  the measured yield distributions 
as a function of Z ( relative to the line of $\protect\beta$ stability,
Z$_{\protect\beta}$ ) and A are presented as contour lines.
The calculated values from DIT/GEMINI are shown as  thick full line
( without acceptance cut ) and  as a thick dashed line ( with acceptance cut ).
The thin dashed line is from the  EPAX parametrization \cite{EPAX}. 
Finally, in Fig. \ref{fig:fig11}c, the velocity vs. mass distributions 
are given.
The present data  are shown as contours as in Fig. \ref{fig:fig11}b. 
The thick full line is from
the DIT/GEMINI calculation without acceptance cut and the dashed line is with 
acceptance cut. 
In general, we see that the DIT/GEMINI calculations are able to provide 
a satisfactory quantitative description of the observed gross 
distributions. Also, it does a fair job in predicting the absolute 
values of the production cross sections ( except for the very n-rich isotopes,
as already pointed out ). 

From a practical standpoint,  using the present  cross section results
we can make estimates of rare beam rates from  intense beams at this
energy regime.
Assuming a beam of 100 pnA  $^{86}$Kr at 25 \hbox{A MeV} striking 
a  10 mg/cm$^{2}$
$^{64}$Ni target,  we give two indicative rate estimates for rare beams:
First, for $^{84}$Se ( two-proton removal product, cross section 6 mb )
the rate is $\sim 3.6 \times 10^{5} $ particles/s.
Second, for the more exotic $^{87}$Se ( two-proton removal+three-neutron pickup 
product, cross section $\sim 12 \mu b$ ) the rate is about 800 particles/s.
Such yields of rare isotopes may enable a variety of nuclear structure and
nuclear reaction studies in the Fermi energy regime.

In general,  from the present experimental study and calculations, 
we see  that such reactions, near the  Fermi energy, involving  
extensive nucleon exchange 
between the 
projectile  and the target, can be utilized as  an efficient way to produce   
very  neutron-rich nuclei.
Apart from in-flight possibilities,
the option of exploiting this type of reaction 
( in normal or inverse kinematics )
at these ( or lower ) energies for rare isotope production in an IGISOL-type 
concept is currently under way  at Texas A\&M.

\section{Conclusions and outlook \label{sec:concl}}

As shown in previous sections, detailed studies of nucleus-nucleus collisions 
in the Fermi energy domain can provide deep insight into multifragmentation 
phenomena. Of special interest are further detailed studies of the isospin 
asymmetry of the emitted fragments which appears to be a sensitive 
probe into the reaction dynamics. 
From a practical point of view, the Fermi energy domain offers unique 
possibilities for production of neutron-rich rare nuclides. 
The ways to utilize such possibilities will be explored in the 
near future. 

\section*{Acknowledgments}

This work was supported in part by the Robert A. Welch Foundation 
through grant No. A-1266 and the Department of Energy through grant 
No. DE-FG03-93ER40773. M. V. was partially supported through 
grant VEGA-2/1132/21.

\end{document}